\RequirePackage{fix-cm}
\documentclass[12pt]{article}                     
\usepackage{graphicx}
\usepackage{hyperref}
\usepackage{amsmath}
\usepackage{amsthm}
\usepackage{amsfonts}
\usepackage{hyperref}
\usepackage[top=1in, bottom=1in, left=0.8in, right=0.8in]{geometry}
\numberwithin{equation}{section}
\begin{document}

\title{Quantum behavior of a classical particle subject to a random force
}


\author{Can Gokler\footnote{Harvard University, Cambridge MA, USA}}



\date{\vspace{-6ex}}

\maketitle

\abstract
We give a partial answer to the question whether the Schr\"{o}dinger equation can be derived from the Newtonian mechanics of a particle in a potential subject to a random force. We show that the fluctuations around the classical motion of a one dimensional harmonic oscillator subject to a random force can be described by the Schr\"{o}dinger equation for a period of time depending on the frequency and the energy of the oscillator. We achieve this by deriving the postulates of Nelson's stochastic formulation of quantum mechanics for a random force depending on a small parameter. We show that the same result applies to small potential perturbations around the harmonic oscillator as long as the total potential preserves the periodicity of motion with a small shift in frequency. We also show that the noise spectrum can be chosen to obtain the result for all oscillator frequencies for fixed mass. We discuss heuristics to generalize the result for a particle in one dimension in a potential where the motion can be described using action-angle variables.

\section{Introduction}

\indent

Despite the successes of quantum theory there remains the solution of the measurement problem and its unification with general relativity. Much effort has been spent assuming quantum mechanics is fundamental and applies to smallest and largest possible scales. At the smallest scales where quantum effects in gravity should take place, although we can form mathematically consistent quantum gravity theories, we have no experimental guidance yet and have extreme conceptual difficulties making sense of a quantum description of space-time. At large scales we observe that nature behaves classically which is not possible to understand within the standard postulates of quantum mechanics since those do no pretense to explain the measurement processes and the quantum to classical transition in a fundamental fashion. Perhaps the best way we know today to solve the measurement problem is to introduce ad hoc spontaneous collapse theories. 

Regardless of the successes and failures of quantum mechanics, in this paper we would like to retain the Newtonian-Einsteinian notions. We try to answer the simplest possible question: Can a single non-relativistic quantum particle in a potential in one dimension can be described by Newtonian mechanics? A lot of effort has been put in deriving hidden variable theories but the answer to this question is still missing: there is no proof that it is impossible and there is no proof that all quantum effects described by a general solution of the Schr\"{o}dinger equation can be accounted classically. Indeed the latter seems almost impossible since it is very hard to imagine how a classical particle would exhibit quantum interference. We do not dogmatically believe that nature is fundamentally classical or quantum but find that it is important to explore the boundaries of existing theories. The least we can expect by answering such a question is a reformulation of quantum theory irrespective of whether the Newtonian description is physically fundamental or not. 

Here we set aside the question of many particles which would involve Bell's theorem. It is widely believed that it is impossible to have a local Newtonian-Ensteinian explanation of entangled states. Although the mostly forgotten rigorous analysis of Bell inequalities by Nelson\cite{Nelson86, Nelson88, Schulz} distinguishes between passive and active locality and makes it possible for stochastic field theories to be in principle able to explain entangled states. We believe that even the explanation of superposition states for a single particle presents an enormous challenge and we do not yet worry about many particle states.  

In this paper we are not able to answer the general question nor we can explain superposition states but we give a partial answer. We show that when a classical harmonic oscillator is subject to a specific random force, its fluctuating motion around the classical trajectory can be described by the Schr\"{o}dinger equation for a range of values of frequency, energy and time. We show that this generalizes to potential perturbations around the harmonic oscillator as long as the periodic motion is preserved with a small shift in frequency. We also show that the noise spectrum can be chosen to obtain the result for all oscillator frequencies for fixed mass. We further discuss a way to generalize the result to potentials admitting action-angle variables. The mathematical tools that we use to achieve this are the method of stochastic averaging and Nelson's formulation of Schr\"{o}dinger equation in terms of stochastic particle trajectories. 

Before the significant discovery of Nelson\cite{Nelson66} that it is possible to give a stochastic account of Schr\"{o}dinger equation, it had been widely believed this was impossible because diffusions are dissipative but there is a notion of conserved energy in the quantum mechanical evolution. Nelson showed that it is possible to construct conservative diffusions which are equivalent to the Schr\"{o}dinger equation. However Nelson's formulation is not Newtonian: the particle is subject to random motion in its position space contrary to that the random effects should appear as forces in a Newtonian theory. Here we attempt to answer whether Nelson's formulation can be derived from a phase space stochastic process where the random term appears as a force. Indeed this is the first and perhaps the most important of the open problems stated in his book\cite{Nelson2}. This question was most openly investigated by Smolin\cite{Smolin} who gave sufficient conditions for a cosmological theory to reduce to Nelson's theory.
 
The same type of questions have been asked and were tried to be answered mostly by the stochastic electrodynamics community\cite{Pena}. There one assumes that an electrically charged particle is coupled to a background stochastic electric field with a specific spectrum and is also subject to electromagnetic radiation reaction. One is able to show that in equilibrium one can choose the spectrum to match with all the energy eigenstates of a harmonic oscillator. However there lacks a universal spectrum working for all energy eigenstates and the superposition states seem to be elusive. There are two main lines of attempted derivations of Schr\"{o}dinger equation both running into difficulties. In the first approach by integrating out the velocity evolution one tries to reduce to a position space process. Schr\"{o}dinger equation holds if one can neglect certain radiative terms in the equations but there is no justification for how the system reaches a state such that those terms can be neglected and how long the system stays in that state such that the approximation is valid. In the second approach it is shown that if one assumes that there are multiple ergodic energy states then stochastic variables can be described by matrix variables and one obtains Heisenberg's theory. However it seems very difficult to construct a stochastic system exhibiting classical multiple ergodic energy states which matches with the quantum energy eigenvalues and to describe the transition between energy eigenstates in such a framework. Perhaps the most important objection against stochastic electrodynamics is that it only applies to charged particles and lacks universality. However we think the questions asked and attempted to be answered in this model are valuable and give insights for further developments. 

The paper is organized as follows. In Section 2 we briefly review Ito calculus and stochastic differential equations-the mathematical framework that we use in the rest of the paper. In Section 3 we give an account of Nelson's stochastic formulation of the Schr\"{o}dinger equation for a non-relativistic particle in one dimension. We introduce the two postulates of Nelson which are equivalent to the Schr\"{o}dinger equation in Madelung form.  In Section 4 we show that the Newton-Nelson law is satisfied by a particle subject to a random force proportional to white noise. In Section 5 we introduce the method of stochastic averaging to be used in the following section to derive Nelson's first postulate. In Section 6 we show that Nelson's first postulate is satisfied for a time interval depending on the energy and the frequency of the oscillator by choosing a suitable spectrum for random force. We further show how a colored spectrum yields Nelson's two postulates for oscillators of all frequencies with fixed mass. We discuss how this result generalizes to small potential perturbations around the harmonic oscillator as long as the periodic motion is preserved with a small shift in frequency. In Section 7 we give heuristics to generalize the result to arbitrary potentials admitting action angle variables. In Section 8 we discuss the results.

\section{Review of stochastic differential equations }

\indent

We give a brief review of Ito stochastic calculus and stochastic differential equations. We will only state results formally which are relevant for our purposes and refer the reader to standard textbooks on the subject(e.g. \cite{Friedman, Arnolds}). Let $\xi(t)$ be the Gaussian process with zero mean and unit variance (also known as white noise), i.e. 

\begin{equation}
\langle \xi(t) \rangle = 0, \; \;  \;\forall t
\end{equation}
and for times $(t_1, t_2, ..., t_n)$, $(\xi(t_1), \xi(t_2), ..., \xi(t_n))$ are Gaussian correlated random variables with co-variance

\begin{equation}
\langle \xi(t_1) \xi(t_2) \rangle = \delta(t_1 - t_2).
\end{equation}
Note that for $t_1 \neq t_2$, $\xi(t_1)$ and $\xi(t_2)$ are independent. We define the Wiener process $W(t)$ as the formal time integral of $\xi(t)$: 

\begin{equation}
W(t) = \int_0^t \xi(s) ds
\end{equation}
where we set the initial time to $t=0$ without loss of generality. We can also write this as $dW(t) = \xi(t) dt$. The Wiener process is again Gaussian since it is a linear combination of independent Gaussian random variables. Its mean is zero as can be directly seen from the definition. Its co-variance is calculated as

\begin{equation}
\langle dW(t_1) dW(t_2) \rangle = \int_0^{t_1} \int_0^{t_2} \langle \xi(s_1) \xi(s_2)\rangle ds_1 ds_2 = \text{min} ( t_1, t_2 ).
\end{equation}
From this we see that formally $dW(t)$ is of order $\sqrt{dt}$. We will be dealing with stochastic differential equations in the rest of the paper. Suppose we would like to make sense of the following initial value problem for the scalar variable $x(t)$:

\begin{equation}
\frac{dx(t)}{dt} = f(x(t)) + g(x(t)) \xi(t)
\end{equation}
with $p(x, t=0))=p_0(x)$ for some initial probability distribution $p_0(x)$. An ambiguity arises when we would like to make sense of the product $g(x(t)) \xi(t)$. We know that since $\xi(t)$ is independent of $\xi(s)$ for $s < t$,  it is independent of $g(x(s))$ for $s < t$. But the product concerns the same times. In order to remedy this difficulty we will write the equation in differential form:

\begin{equation}
dx(t) = f(x(t))dt + g(x(t)) dW(t)
\end{equation}
which is a formal way to write the integral equation:

\begin{equation}
x(t) = x_0 + \int_0^t f(x(s)) ds + \int_0^t g(x(s)) dW(s).
\end{equation}
Now if we can make sense of the integral that includes $dW(s)$ term we can define the stochastic differential equation in terms of the integral equation. There are more than one ways to define a stochastic integral. In this paper we will operate with the Ito definition. For the other famous (Stratonovich) definition see \cite{Friedman, Arnolds}. We adopt the following definition:

\begin{equation}
\int_0^t g(x(s)) dW(s) = \lim_{\Delta s \to 0} \sum_i g(x(s_i)) ( dW(s_{i+1}) - dW(s_{i}))
\end{equation}
where $\Delta s = s_{i+1}-s_i, \; \; \forall i$. Therefore the increment $dW(s_{i+1}) - dW(s_{i})$ is independent of $g(x(s_i))$. However with this definition we need to update the chain rule of calculus. Suppose that we would like to calculate the equation that is obeyed by a function of $x$, say $y=f(x)$. Remember that $dW(t)$ is of order $\sqrt{dt}$. Thus in order to correctly calculate $dy$ we should expand it up to second order. Without proof we state the Ito's lemma:

\begin{equation}
dy = \frac{df}{dx} dx + \frac{1}{2}\frac{d^2 f}{dx^2} (dx)^2 =  \frac{df}{dx} dx +  \frac{1}{2}\frac{d^2 f}{dx^2} g^2(x) dt.
\end{equation}
Note that in the expansion of $(dx)^2$ we omitted terms of order $dt^{3/2}$ and only kept those of order  $dt$ and $\sqrt{dt}$. We will also need the two dimensional version of this. Suppose we have two processes $x_1$ and $x_2$ with independent Wiener processes $dW_1(t)$ and $dW_2(t)$:

\begin{equation}
dx_i(t) = f_i(x_1, x_2)dt + g_i(x_1, x_2) dW_i(t).
\end{equation}
If $y=f(x_1, x_2)$ then we can write the differential $dy$ as

\begin{equation}
dy = \frac{\partial f}{\partial x_1} dx_1 + \frac{\partial f}{\partial x_2} dx_2 +  \frac{1}{2}(\frac{\partial^2 f}{\partial x_1^2} g_1^2 + \frac{\partial^2 f}{\partial x_2^2} g_2^2 )dt.
\end{equation}
We will frequently invoke these results in the following sections. 

\section{Review of Nelson's stochastic mechanics}

\indent

We give a review of Nelson's stochastic formulation of non-relativistic quantum mechanics in one dimension. For more details see Nelson's original paper\cite{Nelson66}, his two books\cite{Nelson1, Nelson2} and Guerra's review\cite{Guerra}. Consider the Schr\"{o}dinger equation:

\begin{equation}
i \hbar \frac{\partial \psi(x,t)}{\partial t} = (-\frac{\hbar^2}{2m} \frac{\partial^2}{\partial x^2} + U(x)) \psi(x,t).
\end{equation}
Putting $\psi(x,t) = \sqrt{\rho(x,t)} e^{\frac{i}{\hbar} S(x,t)}$ we get the Madelung equations:

\begin{equation}
\frac{\partial \rho}{\partial t} = - \frac{\partial}{\partial x} ( \rho \frac{1}{m} \frac{\partial S}{\partial x} )
\end{equation}

\begin{equation}
\label{Madelung2}
\frac{\partial S}{\partial t} = - \frac{1}{2m} ( \frac{\partial S}{\partial x} )^2 - U(x) + \frac{\hbar^2}{2m} \frac{1}{\sqrt{\rho}} \frac{\partial^2}{\partial x^2} \sqrt{\rho}
\end{equation}
where $\rho(x,t)$ is the probability of finding the particle at $(x,t)$ and $S(x,t)$ is the phase of the wave function. We recognize first of the equations as the continuity equation with velocity $\frac{1}{m} \frac{\partial S}{\partial x}$. The second of the equations apart from the last term (quantum potential) on the right hand side is the Hamilton-Jacobi equation. Thus if $\hbar = 0$, we have the classical ensemble of particles. The Newton's equations of motion are then the equations that characteristic curves obey corresponding to this set of Madelung partial differential equations. Since the quantum potential term depends on the probability $\rho(x,t)$, giving deterministic characteristics seems not possible. However as Nelson proved\cite{Nelson66, Nelson1, Nelson2}, it is possible to give a Markovian stochastic process associated to the solution of Madelung equations in position space. We start by assuming that a particle obeys the following stochastic differential equation:

\begin{equation} \label{Nelsonfirst}
 dx(t) = b(x(t), t) dt + \sqrt{\frac{\hbar}{m}} dW(t)
\end{equation}
where $b(x(t),t)$ is a general function and $dW$ is the Wiener process. We call this as Nelson's first postulate. The diffusion equation associated to this is\cite{Friedman, Arnolds}

\begin{equation}
\frac{\partial \rho(x,t)}{\partial t} = - \frac{\partial}{\partial x} ( b(x,t) \rho(x,t) ) +  \frac{\hbar}{2m} \frac{\partial^2}{\partial x^2}\rho(x,t)
\end{equation}
where $\rho(x,t)$ is the probability of finding the particle at $x$ at time $t$. In order to match with the continuity equation we define 

\begin{equation}
\frac{\partial}{\partial x}  S(x,t)= m ( b(x,t) - \frac{\hbar}{2m} \frac{\partial}{\partial x } \log \rho(x,t) )
\end{equation}
where we assumed that $\rho(x,t)$ is nowhere zero. For a discussion of what happens at zeros see \cite{Nelson2}. We want $S(x,t)$ just defined in this way to satisfy the quantum Hamilton-Jacobi equation. We could postulate it as a partial differential equation but Nelson found a way to write this solely in terms of the stochastic particle trajectory. The quantum Hamilton-Jacobi equation can be shown to be equivalent to the following equation:

\begin{equation}
\frac{1}{2}(D_+D_- + D_- D_+ ) x(t) = -\frac{1}{m} \frac{\partial U(x)}{\partial x} |_{x(t)}
\end{equation}
where $D_+$ and $D_-$ are forward and backward derivatives which will be defined below, the right hand side is the classical acceleration of the particle  evaluated on the stochastic trajectory and the left hand side is the time-symmetric stochastic acceleration. This is the stochastic analogue of Newton's second law. Thus we call this as Newton-Nelson law or Nelson's second postulate. The forward and backward derivatives are defined to be

\begin{equation}
D_+ x(t) = \lim_{\Delta t \to 0^+}  E [\frac{x(t+ \Delta t) - x(t)}{\Delta t} | x(t)]
\end{equation}

\begin{equation}
D_- x(t) =\lim_{\Delta t \to 0^+}  E [\frac{x(t) - x(t - \Delta t)}{\Delta t} | x(t)]
\end{equation}
where $E[f| x(t) ]$ denotes the expectation of $f$ conditioned on $x(t)$. For any function $F(x,t)$ we can write its forward and backward derivatives explicitly as follows

\begin{equation} \label{D+}
(D_+F)(x,t) = \frac{\partial}{\partial t} F(x,t) + b(x,t) \frac{\partial}{\partial x} F(x,t) + \frac{\hbar}{2m} \frac{\partial^2}{\partial x^2} F(x,t)
\end{equation}

\begin{equation} \label{D-}
(D_-F)(x,t) = \frac{\partial}{\partial t} F(x,t) + (b(x,t)-\frac{\hbar}{m} \frac{\partial}{\partial x} \log \rho(x,t)) \frac{\partial}{\partial x} F(x,t) - \frac{\hbar}{2m} \frac{\partial^2}{\partial x^2} F(x,t).
\end{equation}
The derivation of the formula for $D_+$ is straightforward but the calculation of $D_-$ is subtler\cite{Nelson1, Nelson2, Guerra}. Using these formulas it is straightforward to show that the Newton-Nelson law is equivalent to the $x$ derivative of the second Madelung equation (equation $\ref{Madelung2}$). It has been shown that for each solution of the Schr\"{o}dinger equation there is an associated stochastic process satisfying Nelson's postulates and if Nelson's postulates are satisfied that one can construct a wave function which satisfies the Schr\"{o}dinger equation with its absolute square the probability density of the position of particle. The stochastic formulation can be generalized to particles propagating in higher dimensions, multiple particles, fields and particles with spin\cite{Nelson2, Guerra}.

\section{Newton-Nelson law}

\indent

In this section we will show that the Newton-Nelson law is satisfied by a particle in a potential in one dimension subject to a random force. Consider a particle of mass $m$ in a potential $U(x)$ subject to a random force:

\begin{align} \label{process}
& dx(t)= v(t) dt \\ \nonumber
& dv(t)= a(x(t))dt+\sigma dW(t)
\end{align}
where $(x, v)$ denotes the position and velocity variables, $dW$ is the Wiener process, $\sigma$ is a positive constant and

\begin{equation}
a(x)= -\frac{U'(x)}{m}=-\frac{1}{m}\frac{dU(x)}{dx}.
\end{equation} 
 We will make use of the following formulas for forward and backward derivatives conditioned on fixed $(x(t), v(t) )$ of a function $G(x,v,t)$ which can be found in section 5 of Guerra's review\cite{Guerra}:

\begin{align} \label{D+phase}
(D_+ G)(x,v,t) |_{(x(t), v(t))} &= \lim_{\Delta t \to 0^+} E[ \frac{G(x(t+\Delta t), v(t + \Delta t), t+\Delta t) - G(x(t), v(t), t)}{\Delta t} | x(t), v(t)]
\\ \nonumber &= \frac{\partial G}{\partial t} + v \frac{\partial G}{\partial x} + a(x) \frac{\partial G}{\partial v} + \frac{\sigma^2}{2} \frac{\partial^2G}{\partial v^2}
\end{align}

\begin{align} \label{D-phase}
 (D_- G)(x,v,t) |_{(x(t), v(t))}  &= \lim_{\Delta t \to 0^+} E[ \frac{G(x(t), v(t), t) - G(x(t-\Delta t), v(t - \Delta t), t-\Delta t) }{\Delta t} | x(t), v(t)] \\ \nonumber &= \frac{\partial G}{\partial t} + v \frac{\partial G}{\partial x} + ( a(x) - \sigma^2 \frac{\partial}{\partial v} \log \rho(x,v,t) )  \frac{\partial G}{\partial v} - \frac{\sigma^2}{2} \frac{\partial^2G}{\partial v^2}
\end{align}
where $\rho(x,v,t)$ is the probability of finding the particle at $x$ with velocity $v$ at time $t$. We also need the following result on conditional expectations for a set of random variables $(x,y,z)$:

\begin{equation} \label{expectation}
E [ F(z) | x] = \int E [F(z) | x, v] p (v| x) dv
\end{equation}
for any function $F(z)$. To derive Newton-Nelson law we will calculate the stochastic acceleration $\frac{1}{2}(D_+D_- + D_- D_+ ) x$. From equations $\ref{D+}$ and $\ref{D-}$ we see that 

\begin{equation}
D_+ x(t) = D_- x(t) = v(t)
\end{equation}
where conditioning on $v(t)$ does not matter. Next we calculate $D_+ D_- x(t)$ and $D_- D_+ x(t)$ conditioned on $(x(t), v(t))$ using equations $\ref{D+phase}$ and $\ref{D-phase}$:

\begin{equation}
D_+ D_- x(t) = D_+ |_{(x(t), v(t))} v(t) = a(x(t))  
\end{equation}

\begin{equation}
D_- D_+ x(t) = D_- |_{(x(t), v(t))} v(t) = a(x(t)) - \sigma^2 \frac{\partial}{\partial v} \log \rho(x,v,t).
\end{equation}
Hence

\begin{equation}
\frac{1}{2} ( D_+ D_- + D_- D_+) x(t)  |_{(x(t), v(t))}  = a(x(t)) - \frac{\sigma^2}{2} \frac{\partial}{\partial v} \log \rho(x,v,t).
\end{equation}
In order the calculate the stochastic acceleration, which is conditioned only on $x(t)$, we use equation$\ref{expectation}$:

\begin{align}
\frac{1}{2} ( D_+ D_- + D_- D_+) x(t)  |_{x(t)} &= \int \frac{1}{2} ( D_+ D_- + D_- D_+) x(t)  |_{(x(t), v(t))} p_t(v|x) dv \\ \nonumber &= a(x(t)) -  \frac{\sigma^2}{2} \int \frac{\partial p_t(v | x)}{\partial v} dv = a(x(t)).
\end{align}
Thus we have shown that the Newton-Nelson law is satisfied by the process given by equation $\ref{process}$. This result was stated without proof in \cite{Nelson1} for the particle in a potential subject to linear friction in equilibrium.

\section{Method of stochastic averaging}

\indent

In this section we introduce the method of averaging of stochastic differential equations. There are several formulations of stochastic averaging though we will only consider the theorem due to Khas'minskii\cite{Khas, Strat, Freidlin, Spanos, Pavliotis} applied to two dimensional systems in Ito form. Consider the process $(x,y)$:

\begin{equation}
dx(t) = \epsilon^2 f_1 (x(t), y(t), t) dt + \epsilon g_1(x(t),y(t),t) dW(t)
\end{equation}
\begin{equation}
dy(t) = \epsilon^2 f_2 (x(t), y(t), t) dt + \epsilon g_2(x(t),y(t),t) dW(t)
\end{equation}
where $dW$ is the Wiener process and $0 <\epsilon \ll 1$ which means that $(x,y)$ are slowly varying in time as compared to $f_i$ and $g_i$. We assume that $f_i$ and $g_i$ are sufficiently continuously differentiable and bounded.  Then for times of order $O(1 / \epsilon)$ the dynamics can be uniformly approximated the following averaged system \footnote{More precisely the original process converges weakly to the averaged process as $\epsilon \rightarrow 0$.}:

\begin{equation}
dx(t) = \epsilon^2 \bar{f}_1(x(t),y(t)) dt + \epsilon \sigma_{11}(x(t),y(t)) dW_1(t) + \epsilon \sigma_{12}(x(t),y(t)) dW_2(t)
\end{equation}

\begin{equation}
dy (t)= \epsilon^2 \bar{f}_2(x(t),y(t)) dt + \epsilon \sigma_{21}(x(t),y(t)) dW_1(t) + \epsilon \sigma_{22}(x(t),y(t)) dW_2(t)
\end{equation}
where $dW_1$ and $dW_2$ are independent Wiener processes and the averaged functions are given by

\begin{equation}
\bar{f}_i(x,y) = \lim_{T \to \infty} \frac{1}{T} \int_0^T  f_i(x,y,t) dt 
\end{equation}

\begin{equation}
(\sigma \sigma^T)_{ij} (x,y) = \lim_{T \to \infty} \frac{1}{T} \int_0^T g_i(x,y,t) g_j(x,y,t) dt
\end{equation}
where $\sigma^T$ denotes the matrix transpose of $\sigma$. Note that $\sigma$ is unique up to a time dependent orthogonal transformation $R(t)$ as $\sigma R(t) (\sigma R(t))^T=\sigma \sigma^T$. It can be shown that $R(t) [dW_1(t) \: dW_2(t)]^T$ is again a Wiener process therefore replacing $\sigma $ with $\sigma R(t)$ does not change the diffusion process $(x,y)$. For applications below we need the periodic version of averaging. For periodic systems we can write

\begin{equation}
\bar{f}_i(x,y) =  \frac{1}{T} \int_0^T  f_i(x,y,t) dt 
\end{equation}

\begin{equation} \label{ave}
(\sigma \sigma^T)_{ij} (x,y)  = \frac{1}{T} \int_0^T g_i(x,y,t) g_j(x,y,t)dt
\end{equation}
where $T$ is the period of oscillations which correspond to $dx=dy=0$. The way that $dx = dy =0$ corresponds to a periodic deterministic solution will be clarified in the examples in the following sections. The stochastic averaging principle is a generalization of its deterministic version which can found in \cite{Sanders}. For deterministic averaging of a one dimensional system in action-angle variables see \cite{Arnold}. For more on stochastic averaging see the review\cite{Spanos} and the books \cite{Freidlin, Pavliotis}.

\section{Nelson's first postulate for a harmonic oscillator}

\indent

Consider the harmonic oscillator with frequency $\omega$ and mass $m$ subject to a random force with position and velocity variables $(x,v)$:

\begin{align} \label{osc}
& dx(t)=v(t)dt \\ \nonumber
& dv(t)=-\omega^2x(t)dt+\epsilon \omega dW(t).
\end{align}
Assume that the initial energy of the oscillator is $E_0$ is probability 1. We will show that we can make the choice $\epsilon=\sqrt{\frac{2\hbar}{m}}$ such that the position process $x(t)$ is approximately Markovian and satisfies Nelson's first postulate for a time interval depending on $\hbar$, $E_0$ and $\omega$. We show this by approximating the dynamics given by equation $\ref{osc}$ by an averaged process using the method of stochastic averaging. We will see that this will induce a noise term for the $x$ variable which is necessary to satisfy Nelson's first postulate. Note that the noise term appearing in equation $\ref{osc}$ is not necessarily small as it is proportional to $\epsilon \omega$ as $\omega$ can be large.

To proceed note that the dynamics in phase space is not in standard form for averaging. Therefore apply the coordinate transformation 

\begin{align} \label{defx}
& x=r\cos(\omega t + \phi) \\ \nonumber
& v=-\omega r \sin(\omega t+ \phi)
\end{align}
or

\begin{align}
& r=\sqrt{x^2+\frac{v^2}{\omega^2}} \\ \nonumber
& \phi=-\arctan(\frac{v}{\omega x})-\omega t.
\end{align}
To calculate the differential of $r$ and $\phi$ we use Ito's lemma and obtain

\begin{align}
& dr=\frac{x}{r}dx+\frac{v}{\omega^2 r}dv+\frac{1}{2}\frac{x^2}{ \omega^2 r^3} (dv)^2=\frac{(\epsilon\omega)^2}{2}\frac{x^2}{ \omega^2 r^3}dt+\epsilon \omega \frac{v}{\omega^2 r} dW \\ \nonumber
& d\phi=\frac{v}{\omega r^2}dx-\frac{x}{\omega r^2}dv + \frac{1}{2} \frac{2xv}{\omega^3r^4}(dv)^2-\omega dt = (\epsilon \omega)^2 \frac{xv}{\omega^3r^4}dt - \epsilon \omega \frac{x}{\omega r^2 }dW.
\end{align}
We see that both $r$ and $\phi$ are slowly varying. Therefore we apply the method of averaging over one period $T=\frac{2\pi}{\omega}$ of the harmonic oscillator. This amounts to fixing $r$ and averaging over the angle variable. Denote the time average of a function $f(x,v)$ by

\begin{equation}
 \langle f(x,v) \rangle_T=\frac{1}{T} \int^T_0 f(x(t),v(t))dt.
 \end{equation}
The evolution equations can be approximated by the following averaged equations over time intervals of order $O(1/\epsilon)$ noting that the noise terms are of order $\epsilon$:

\begin{align}
& dr=\frac{(\epsilon\omega)^2}{2}\frac{1}{ \omega^2 r^3} \langle x^2 \rangle_T dt+\epsilon \omega \frac{1}{\omega^2 r} \sqrt{\langle v^2 \rangle_T} dW_1  \\ \nonumber
& d\phi= (\epsilon \omega)^2 \frac{1}{\omega^3r^4} \langle xv\rangle_Tdt + \epsilon \omega \frac{1}{\omega r^2 } \sqrt{\langle x^2 \rangle_T} dW_2
\end{align}
where $dW_1$ and $dW_2$ are independent Wiener processes. The averaged quantities are calculated to be

\begin{align}
&\langle x^2 \rangle_T= \frac{1}{2 \pi} \int_0^{2 \pi} r^2 \cos^2\theta d\theta = \frac{r^2}{2} \\ \nonumber
&\langle v^2 \rangle_T = \frac{\omega^2}{2 \pi} \int_0^{2 \pi} r^2 \sin^2\theta d\theta = \frac{\omega^2 r^2}{2} \\ \nonumber
&\langle xv \rangle_T = -\frac{\omega}{2 \pi} \int_0^{2 \pi} r^2 \sin\theta \cos \theta d\theta = 0
\end{align}
where we have chosen $\sigma$ in equation $\ref{ave}$ as diagonal as $\langle xv \rangle_T=0 $. Substitute the averaged quantities in the averaged equations to get:

\begin{align} \label{avr}
& dr=\frac{\epsilon^2}{4}\frac{1}{  r} dt+\epsilon \frac{1}{\sqrt{2}} dW_1  \\ \nonumber
& d\phi=  \epsilon \frac{1}{\sqrt{2} r }  dW_2.
\end{align}
The averaged evolution of the amplitude of oscillations $r$ is independent of $\phi$ and the evolution of $\phi$ is determined by the evolution of $r$. Using averaged equations we can derive the averaged evolution of the position variable $x$ using equations $\ref{defx}$ and $\ref{avr}$:

\begin{align}
dx & =dr \cos(\omega t + \phi ) - r \sin(\omega t +\phi) (\omega dt + d\phi)-\frac{1}{2} r \cos(\omega t + \phi) (d\phi)^2 \\ \nonumber
& = v dt + \epsilon \frac{1}{\sqrt{2} r }(xdW_1+\frac{v}{\omega}dW_2).
\end{align}
Note that the method of stochastic averaging induced a stochastic term for the $x$ variable where in the original dynamics defined by equation $\ref{osc}$ the stochastic term only appears in the dynamics of the velocity variable. We can simplify the stochastic term noting that given $x(t)$ and $v(t)$, $dW_1(t)$ and $dW_2(t)$ are independent Gaussian processes. A linear combination  

\begin{equation}
a(x(t),v(t))dW_1(t)+b(x(t),v(t))dW_2(t)
\end{equation} 
of independent zero mean Gaussian processes is again a zero mean Gaussian process with variance $a^2(x(t),v(t))+b^2(x(t),v(t))$. Therefore the equation for the position variable can be written as:

\begin{equation}
dx = v dt + \frac{\epsilon }{\sqrt{2}}dW
\end{equation}
where $dW$ is the Wiener process. In general this is not a Markov process since $v$ itself is fluctuating and is dependent on $x$. However if somehow we can assume that the amplitude $r$ is constant then we can express $v$ in terms of $x$ as 

\begin{equation}
v=\pm \omega \sqrt{r^2-x^2}
\end{equation}
obtaining the Markov process

\begin{equation}
dx = \pm \omega \sqrt{r^2-x^2} dt+ \frac{ \epsilon}{\sqrt{2}}dW.
\end{equation}
Now to match with Nelson's first postulate (equation $\ref{Nelsonfirst}$) we must choose 

\begin{equation}
\epsilon = \sqrt{\frac{2\hbar}{m}}.
\end{equation}
With this choice of $\epsilon$ we can justify the assumption that $r$ remains approximately constant as follows. Assume that initially $r=r_0$ with probability 1. For sufficiently small times we can assume that $r$ is well approximated by $r_0$. To see this introduce the energy variable 

\begin{equation}
E=\frac{1}{2}m\omega^2r^2.
\end{equation}
Using Ito's lemma its dynamics is calculated to be:

\begin{equation}
dE = (\epsilon \omega)^2 \frac{m}{2} dt + \epsilon \omega\sqrt{m E} dW_1
\end{equation}
with $E_0=\frac{1}{2} m \omega^2 r_0^2$. Set  $E(t)=E_0 + \delta E(t)$. Roughly $\delta E(t)$ grows as 

\begin{equation}
\text{max}(\epsilon^2 \omega^2 m t, \epsilon \omega \sqrt{m E_0 t} )
\end{equation}
where $\sqrt{t}$ dependence arises from the Wiener term. The method of stochastic averaging is accurate for times $O(1 / \epsilon)$. Therefore if $1 / \epsilon \ll  1 / \hbar \omega^2$ and $1 / \epsilon \ll 1 / \hbar \omega^2  E_0$ which are satisfied for sufficiently small $\omega$ and $E_0$, we can assume that $\delta E(t)$ is small therefore $E(t) \approx E_0$ and $r(t) \approx r_0$ over the time interval $1 / \epsilon$ in which the method of stochastic averaging is accurate. We can write the dynamics in the following suggestive form

\begin{equation}
dx = v(x, E_0) dt + \sqrt{\frac{\hbar}{m}} dW
\end{equation}
where $v(x, E_0)=\pm \sqrt{\frac{2}{m} (E_0-\frac{1}{2}m\omega^2x^2)}$ is the classical velocity of the particle with energy $E_0$. If the stochastic term is absent then this equation would be the classical equation of motion for the particle. Hence the phase space diffusion process gives rise to a position space Markov process as a small random fluctuation around the classical trajectory.

We initially made the assumption that the random force depends on the frequency of the oscillator. Then it is natural to ask whether we can choose a universal random force term which would give the same result for an arbitrary frequency. Such a choice is indeed possible. So instead of the Markovian model start from 

\begin{align} \label{osc2}
&\dot{x} (t) = v (t)\\ \nonumber
&\dot{v} (t) = -\omega^2 x(t) + \xi(t).
\end{align}
Let $\xi(t)$ be a zero mean Gaussian process with covariance 

\begin{equation}
\langle \xi(t) \xi(t+\tau) \rangle = c(\tau) 
\end{equation}
with its Fourier transform, the power spectrum 

\begin{equation}
S(\Omega)=\int_{-\infty}^{\infty} c(\tau) e^{-i\Omega \tau} d\tau.
\end{equation}
It can be shown that upon averaging the $(r, \phi)$ evolution over one period of the oscillator, only the resonant term corresponding to $\Omega=\omega$ contributes to the averaged equations \cite{Spanos}: 

\begin{align}
& dr=\frac{ S(\omega)}{8 \pi}\frac{1}{ \omega^2 r} dt+  \sqrt{ \frac{S(\omega)}{4 \pi}} \frac{1}{\omega}  dW_1  \\ \nonumber
& d\phi=  \sqrt{ \frac{S(\omega)}{4 \pi}}  \frac{1}{\omega r }  dW_2.
\end{align}
Thus if we choose $S(\Omega)=\frac{4 \pi \hbar}{m}\Omega^2$, we recover the previous results. Note that the averaged equations obtained for the colored noise are the same as the ones obtained before for white noise with spectrum dependent on frequency. Hence the processes defined by equations $\ref{osc}$ and $\ref{osc2}$ are good approximations to each other. Therefore Newton-Nelson law is approximately satisfied for the colored noise case as it is satisfied for equation  $\ref{osc}$.

Now we consider small potential perturbations around the harmonic oscillator:

\begin{align}
& dx(t)=v(t)dt \\ \nonumber
& dv(t)=-\omega^2x(t)dt -  \eta \frac{1}{m}\frac{d U(x)}{dx}|_{x=x(t)}+\epsilon \omega dW(t)
\end{align}
where $\frac{d U(x)}{dx}$ is $O(1)$ and $\eta \ll 1$. We assume that $U(x)$ preserves the periodic structure where the new frequency $\tilde{\omega}(E)$ is a small perturbation of the frequency of the harmonic oscillator:

\begin{equation}
\tilde{\omega} (E) = \omega + \eta \delta \omega.
\end{equation}
Since $\eta \delta \omega$ induces a $O(\epsilon \eta)$ correction in the stochastic term, after repeating the steps for the pure harmonic oscillator, we can approximately write 

\begin{equation}
dq = v(x, E_0) dt + \frac{\epsilon}{\sqrt{2}} dW
\end{equation}
where this time $v(q, E_0)$ is the velocity associated to the perturbed potential: 

\begin{equation}
v(x, E_0) = \pm \sqrt{ \frac{2}{m}( E_0 - \frac{1}{2}m\omega^2 x^2 - \eta U(x)) }.
\end{equation}

\section{Nelson's first postulate for a general potential}

\indent

In this section we would like to generalize the results for the harmonic oscillator to a general potential. However we will be able to show less. Due to the difficulty in calculating averages explicitly for general potentials we will be only able to show that we can choose the random force dependent on energy (unlike the harmonic oscillator case where the random force is independent of coordinates) such that Nelson's first law is satisfied using a heuristic averaging procedure. We will restrict to potentials for which the motion can be described using action-angle variables. Therefore consider a particle of mass $m$ in one dimension in a potential $U(x)$ subject to a random force:

\begin{align}
& dx(t)=\frac{p(t)}{m} dt \\ \nonumber
& dp(t)=-U'(x(t)) dt+\epsilon dW(t)
\end{align}
where $(x,p=mv)$ denotes the position and momentum variables and $U'(x)=\frac{dU(x)}{dx}$. We first perform the coordinate transformation from $(x, p)$ to $(x, E)$ where $E$ is the energy of the particle defined by

\begin{equation}
E (x,p) = \frac{p^2}{2m}+U(x).
\end{equation}
Using Ito's lemma we calculate $dE$ as

\begin{equation}
dE = U'(x)dx+\frac{p}{m}dp+\frac{1}{2m}(dp)^2=\frac{\epsilon^2}{2m}dt + \epsilon \frac{p(x,E)}{m} dW.
\end{equation}
Note that if the stochastic term is absent then the energy would be conserved. We can also express $dx$ in terms of $(x,E)$ by solving for $p$ in terms of $(x,E)$ in the definition of energy:

\begin{equation}
dx = \pm \frac{1}{m}\sqrt{2m(E - U(x))}dt.
\end{equation}
We now assume that the classical motion can be described by action-angle variables $(\phi, I)$\cite{Arnold}. In terms of the action-angle variables the classical deterministic equations of motion can be written as 

\begin{align}
& d\phi = \omega(I) dt \\ \nonumber
& dI = 0
\end{align}
where the frequency is 

\begin{equation}
w(I)=\frac{dE(I)}{dI}
\end{equation}
and the energy is a function of the action variable alone. Instead of the action variable we will use the energy variable since the energy is a function of the action but not the angle variable. We further assume that this mapping is one-to-one. Define the action function (not the action variable) as

\begin{equation}
S(I, x) = \int_{x_0}^x p(x', E) dx'
\end{equation}
for an arbitrary initial point $x_0$. Then the action variable is defined to be proportional to the action function $S$ over one period of motion:

\begin{equation}
I = \frac{1}{2 \pi} \oint p dx'
\end{equation}
and the angle variable is
\begin{equation}
\phi = \frac{\partial S(I, x)}{\partial I} = \int_{x_0}^x \frac{\partial p(x',E)}{\partial I} dx'.
\end{equation}
Taking the derivative inside the integral we have

\begin{equation}
\label{defphi}
\phi= \omega(I)\int_{x_0}^x \frac{\partial p(x',E)}{\partial E} dx' = m \omega(I) \int_{x_0}^x \frac{1}{p}dx'.
\end{equation}
We define 

\begin{equation}
f(x,E)=\int_{x_0}^x \frac{1}{p}dx'
\end{equation}
so that 

\begin{equation}
\phi(x, E)=m \omega(I(E)) f(x,E).
\end{equation}
Having defined the angle variable we are ready to perform the change of coordinates from $(x, E)$ to $(\phi, E)$. Using Ito's lemma we calculate $d\phi$ as

\begin{align}
d\phi &= m \frac{\partial}{\partial E} (\omega f)dE + \frac{m}{2} \frac{\partial^2}{\partial E^2} (\omega f) (dE)^2 + m\omega \frac{\partial f}{\partial x } dx \\ \nonumber
&= \omega dt + \frac{\epsilon^2}{2}(\frac{\partial}{\partial E}(\omega f) + \frac{p^2}{m}\frac{\partial^2}{\partial E^2} (\omega f))dt+\epsilon\frac{\partial}{\partial E}(\omega f) p dW.
\end{align}
We see that $\phi$ is slowly varying except the $\omega dt$ term. In order to have all the right hand side terms small we further introduce the new angle variable $\theta$ as

\begin{equation}
\theta = \phi - \omega t
\end{equation}
and compute its differential as

\begin{align}
d \theta &= d\phi-\frac{d \omega}{dE} t dE - \frac{1}{2} \frac{d^2 \omega}{dE^2} t (dE)^2-\omega dt \\ \nonumber
&= \frac{\epsilon^2}{2}(\frac{\partial}{\partial E}(\omega f) + \frac{p^2}{m}\frac{\partial^2}{\partial E^2} (\omega f)-t \frac{d^2 \omega}{dE^2} \frac{p^2}{m^2}) dt - \frac{\epsilon^2}{2m}\frac{d \omega}{dE}t dt  \\ \nonumber
& \; \; \; \; + \epsilon\frac{p}{m}(m\frac{\partial}{\partial E}(\omega f)-t\frac{d\omega}{dE})dW.
\end{align}
We have finalized the set of coordinate transformations which yielded slowly varying $(E, \theta)$ coordinates. Next we average the dynamics over a period $T=\frac{2\pi}{\omega(E)}$ fixing $E$ in $(x(t),p(t))$ to obtain the approximate averaged equations. The averaged equations for $(E, \theta)$ are

\begin{equation} \nonumber
dE = \frac{\epsilon^2}{2m} dt + \epsilon \sigma_{11}(E)dW_1 + \epsilon \sigma_{12}(E) dW_2
\end{equation}

\begin{equation}
d \theta = \frac{\epsilon^2}{2}F(E) dt+ \epsilon \sigma_{21}(E)dW_1 + \epsilon \sigma_{22}(E) dW_2
\end{equation}
where $dW_1$ and $dW_2$ are independent Wiener processes, $F(E)$ is given by

\begin{equation}
F(E) = \langle \omega \frac{\partial f}{\partial E} + \frac{p^2}{m} (2 \frac{d \omega}{dE} \frac{\partial f}{\partial E} + \omega \frac{\partial^2 f}{\partial E^2} ) \rangle_T
\end{equation}
and the matrix $\sigma$ is determined from 

\begin{equation}
(\sigma \sigma^T)_{ij}= D_{ij}
\end{equation}
The components of $D$ are the averages

\begin{equation} \nonumber
D_{11}=\langle \frac{p^2}{m^2} \rangle_T
\end{equation}

\begin{equation} \nonumber
D_{12}=D_{21}= \langle \frac{p^2}{m} \omega \frac{\partial f}{\partial E} \rangle_T
\end{equation}

\begin{equation}
D_{22}=\langle p^2 \omega^2 (\frac{\partial f}{\partial E})^2 \rangle_T
\end{equation}
where we have simplified $F(E)$ and $D_{ij}$ noting that since when we take averages over the classical trajectory we can set $dx = \frac{p}{m}dt$ so that $t=mf$. Now we will go back to the dynamics of $x$ to check if Nelson's first postulate is satisfied. First calculating the dynamics of $(E, \phi)$ then 
calculating the dynamics of $x$ we obtain the equation of the form





\begin{equation}
 dx = \frac{p}{m} dt + \epsilon^2 K(x,p) + \epsilon (G_1(x,p)dW_1 + G_2(x,p)dW_2 )
 \end{equation} 
for some functions $K(x,p)$, $G_1(x,p)$ and $G_2(x,p)$. The stochastic term is dependent on coordinates. We invoke without rigorous justification a heuristic averaging procedure although $dx$ is not in the standard form. We fix $E$ and average the small terms over the angles. 

\begin{equation}
dx = \frac{p}{m} dt + \epsilon^2 \bar{K}(E) + \epsilon (\bar{G}_1(E) dW_1 + \bar{G}_2(E) dW_2)
\end{equation}
where $\bar{K}(E)$, $\bar{G}_1(E) $ and $\bar{G}_2(E)$ are the averages of $K(x,p)$, $G_1(x,p)$ and $G_2(x,p)$. Since $dW_1(t)$ and $dW_2(t)$ are independent zero mean Gaussian processes we can write

\begin{equation}
dx = \frac{p}{m} dt + \epsilon^2 \bar{K}(E) + \epsilon G(E) dW
\end{equation}
where $dW$ is the Wiener process and $G^2(E)=\bar{G}_1(E)^2+\bar{G}_2(E)^2 $. Now as in the harmonic oscillator case for sufficiently small times $E$ is almost constant.  We are not able to calculate $G(E)$ explicitly in terms of $E$(or the action variable $I$) and $\omega(E)$ and its derivatives since the averages appearing in $D_{ij}$ seem to be difficult to evaluate analytically. Therefore unlike the harmonic oscillator case we have the weaker result: one can choose 

\begin{equation}
\epsilon G(E) = \sqrt{\frac{\hbar}{m}}
\end{equation}
to satisfy Nelson's first postulate. However in this case we need to choose $\epsilon$ dependent on $E$ and this is unfavorable regarding universality: for every value of energy we need to choose a different correlation coefficient for the noise. We saw in the harmonic oscillator case that choosing the colored random force with spectrum proportional to $\omega^2$ we can satisfy Nelson's first postulate for all oscillator frequencies. If one could evaluate $G(E)$ explicitly one can check whether the same random force spectrum gives rise to Nelson's first law for arbitrary potentials admitting action-angle description. 

\section{Scholia}

\indent

We have shown that it is possible to choose a random force such that fluctuations around the classical trajectory can be described by the Schr\"{o}dinger equation for the harmonic oscillator. We achieved this by showing that the Newton-Nelson law is satisfied for general potentials and Nelson's first postulate is approximately satisfied for states described by constant energy upon averaging the dynamics. We further generalized this result to the case of small perturbations around the harmonic oscillator. We also showed that there exists a colored Gaussian noise such that Nelson's two postulates are approximately satisfied for an oscillator with any frequency with fixed mass. We attempted to generalize the results for the harmonic oscillator to a general potential admitting action-angle variables. However due to the difficulty in evaluating averages and the need for an heuristic averaging principle we are only able to show that there is a random force depending on energy, mass and frequency of the system such that we can obtain Nelson's first postulate. 

In this paper we did not touch the issue of superposition states. Although Nelson's formulation is equivalent to the Schr\"{o}dinger equation and describes superposition states there is no guarantee that the Nelson's process associated to a superposition state could be derived from a phase space process. The simplest idea is to let the particle assume two energies with their respective probabilities. Then the position space process should approximate the superposition state of the respective energies. However this does not work since we are not superposing quantum states but rather considering an ensemble of states and we end up in a mixed state. Thus we do not yet know whether superpositions and the most general quantum states can be described within the framework we presented here. 

\bibliographystyle{unsrt}
\bibliography{derivation_of_nelson_references}

\end{document}